\documentclass[twocolumn,preprintnumbers,amsmath,amssymb]{revtex4}

\usepackage{array}
\usepackage{graphicx}
\usepackage{tabularx}
\usepackage{longtable}
\usepackage{dcolumn} 
\usepackage{bm}

\begin{document}

\title{Thermal rectification in bulk materials with asymmetric shape
}

\author{D. Sawaki$^1$, W. Kobayashi$^{*,2,3}$, Y. Moritomo$^2$, and I. Terasaki$^4$}
\affiliation
{$^1$Department of Physics, Waseda University, Tokyo 169-8555, Japan}
\affiliation
{$^2$Graduate School of Pure and Applied Science, University of Tsukuba, Ibaraki 305-8571, Japan}
\affiliation
{$^3$PRESTO, Japan Science and Technology Agency, Saitama 332-0012, Japan}
\affiliation
{$^4$Department of Physics, Nagoya University, Aichi 464-8602, Japan}
\email{kobayashi.wataru.gf@u.tsukuba.ac.jp}

\date{\today}

\begin{abstract}

We investigate thermal rectification in a bulk material with a pyramid shape 
to elucidate shape dependence of the thermal rectification, 
and find that rectifying coefficient $R$ is 1.35 for this shape, 
which is smaller than $R$=1.43 for a rectangular shape. This result is fully duplicated by our 
numerical calculation based on Fourier's law. We also apply this calculation 
to a given shape, and show a possible way to increase $R$ depending on the shape.

\end{abstract}

\maketitle

A thermal rectifier is a device in which heat flows in a forward direction, 
while it hardly flows in the opposite direction. 
Owing to controllability of the heat current, the thermal rectifier can be 
applied to phononic devices as a diode is essential for electronic devices. 
Since the first report on thermal rectification at an interface between Cu and CuO \cite{1st}, 
this research field has been developed by both experimental and theoretical studies \cite{history}. 
In particular, several advanced studies concerning a discovery of thermal rectification in 
non-uniformly-mass-loaded carbon nanotube \cite{nanotube} and a prediction of 
thermal rectification in anharmonic 1D lattice model/quantum thermal systems 
\cite{terraneo, Li1, JPEckmann, massgrade} have been reported in 2000s, which gives renewal interests to 
this research field. After these studies, thermal transistor \cite{transistor}, thermal memory \cite{memory}, 
and thermal logic gate \cite{logicgate} are proposed as possible applications of the 
thermal rectifier. 

\begin{figure}[t]
\begin{center}
\vspace*{0cm}
\includegraphics[width=4.5cm,clip]{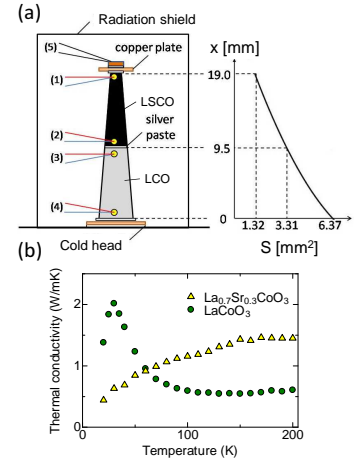}
\caption{(Color online) (a) Schematic figure of the thermal rectifier (LSCO-top configuration). 
(1)-(4) represent chromel-constantan thermocouples, (5) resistive heater. $S(x)$ 
indicates sectional area of the thermal rectifier. (b) Temperature dependence of the 
thermal conductivity of LaCoO$_3$ and La$_{0.7}$Sr$_{0.3}$CoO$_3$ polycrystalline samples.
}
\end{center}
\end{figure}

Recently, we have demonstrated thermal rectification of an oxide thermal 
rectifier as a bulk effect \cite{kobayashi} according to a theory based on Fourier's law \cite{peyrard}. 
This theory predicts thermal rectification in a sample made of two different materials 
with different temperature dependences of thermal conductivities. 
According to Fourier's law, 
\begin{equation}
J=-\kappa [x,T(x)]\frac{dT(x)}{dx},
\end{equation} 
heat flux $J$ is proportional to thermal conductivity $\kappa$. 
Since $\kappa$ is a function of position $x$ and absolute temperature $T(x)$, 
the averaged thermal conductivity $\kappa_{\rm av}$ of the sample may 
change depending on a direction of the temperature gradient 
under finite temperature difference $\Delta T$ (See Fig. 1 of Ref. \cite{kobayashi} or Ref. \cite{peyrard}). 
Thus, the heat flux in a forward direction ($J_{f}$) can 
be different from that in the reverse direction ($J_{r}$) through Eq. (1) 
leading to $R>1$, where $R$ represents the thermal rectifying coefficient defined by a ratio of 
the heat fluxs in the forward and reverse directions 
($R \equiv \frac{\lvert J_f \rvert}{\lvert J_r \rvert}$). 
We have found $R=1.43$ at 40 K with $\Delta T=60$ K in the thermal rectifier 
made of LaCoO$_3$ and La$_{0.7}$Sr$_{0.3}$CoO$_3$ \cite{kobayashi}. 

There are several strategies to achieve larger $R$ in bulk materials as follows. 
(1) To utilize an interface between two materials 
with different phonon bands. In this case, $R$ is predicted to be up to 2000 \cite{li2}. 
(2) To provide materials which exhibits structural phase transition accompanying 
a rapid change of $\kappa $ at the transition temperature. 
(3) To form a thermal rectifier with a proper shape. 
It is theoretically predicted that carbon nanohorn and carbon nanoribon 
exhibit thermal rectification because of larger heat current from a larger section to the 
smaller section \cite{nanocorn,nanocorn2,nanoribon,Hu,topology}. 
This kind of the shape effect is also possible in bulk materials according to Eq. (1). 
However, experimental studies on shape-dependent thermal rectification 
have not been reported yet. In this letter, we will address the issue(3).

\begin{figure}[t]
\begin{center}
\vspace*{0cm}
\includegraphics[width=6cm,clip]{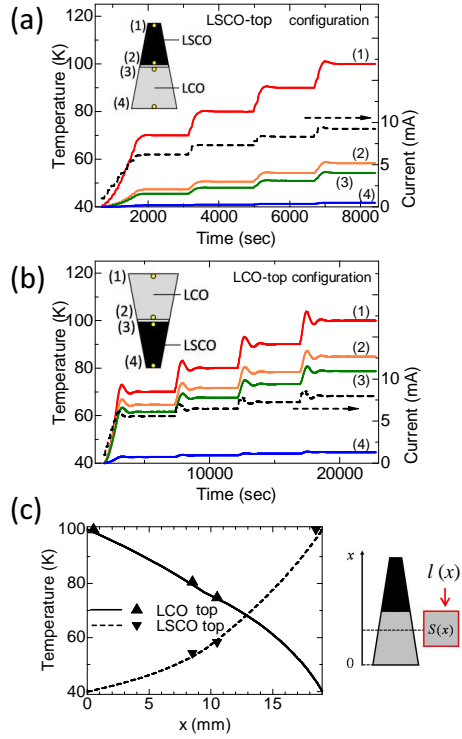}
\caption{(Color online) Monitored temperatures as a function of time at 
points (1)-(4) of (a) LCO-top and (b) LSCO-top configurations, respectively. 
(c) Temperature distributions in the thermal rectifier with LCO-top and LSCO-top configuration, 
respectively. Solid and dotted lines represent numerically calculated temperature distribution 
by Eq. (2). $S(x)$ and $l(x)$ are sectional area and the length of 
periphery of $S(x)$, respectively. 
}
\end{center}
\end{figure}

Polycrystalline samples of LaCoO$_3$ (LCO) and La$_{0.7}$Sr$_{0.3}$CoO$_3$ (LSCO) 
were prepared by a solid state reaction. 
Stoichiometric amounts of La$_2$O$_3$, SrCO$_3$ and Co$_3$O$_4$ were thoroughly mixed, 
and calcined at 1273 K for 24h. 
The products were finely ground, filled up in latex tubes with 
a small amount of water to increase sample's porosity, and pressed at 40 MPa. 
Then, the pressed sample was calcined at 1273 K for 24h. 
X-ray diffraction pattern was measured using 
a standard diffractometer with Fe $K\alpha $ radiation. Any impurity peaks were 
not detected. A mass density of the sample is evaluated to be 50.7$\%$ of the ideal 
density for LCO, and 54.5$\%$ for LSCO, respectively. 

Our measurement system is depicted as a schematic figure in Fig.1 (a). 
First, LCO and LSCO sample bars were cut to let them have a pyramid-like shape. 
The sectional area and length of the samples are shown in the right panel of Fig. 1(a). 
Then, the samples were bonded by silver paste with high thermal conductivity (Kyocera Chemical CT285), 
and annealed at 423 K for 0.5 h and at 493 K for 1.5 h to dry the paste. 
To make heat current $I_{\rm h}$ into the rectifier, a resistive heater with $r=120$ $\Omega$ was 
put on the top of the rectifier together with a copper plate at point(5). 
Another copper plate was attached on the bottom of the 
rectifier and the plate was glued to the cold head of closed cycle refrigerator by GE varnish (GE7031). 
Chromel-constantan thermocouples were attached at points (1)-(4) to monitor temperatures 
in the presence of the heat current. Detailed description of the measurement system is 
given in the previous reports \cite{kobayashi,kobayashi2}. 
Thermal conductivity was measured by a conventional steady state technique using 
chromel-constantan differential thermocouple. 
As shown in Fig. 1(b), $\kappa$ for LCO and LSCO is 5 times smaller than that 
of a single crystal \cite{berggold}, respectively, which is attributed to the high porosity 
of the samples. This low thermal conductivity is important to maintain the large temperature 
difference between points (1) and (4).

\begin{figure}[t]
\begin{center}
\vspace*{0cm}
\includegraphics[width=4cm,clip]{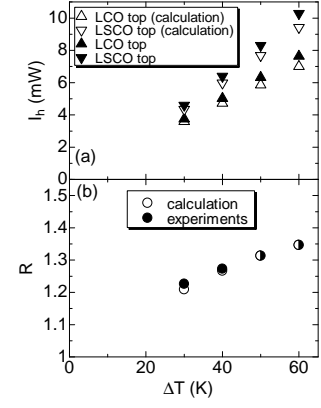}
\caption{(Color online) Calculated and measured (a) heat currents and 
(b) the rectifying coefficient $R$ of both configurations.
}
\end{center}
\end{figure}

Fig. 2(a) and 2(b) show the monitored temperatures as a function of time 
at points (1)-(4) in the forward direction (LSCO-top configuration) 
and the reverse direction (LCO-top configuration), respectively. 
With increasing electric current $I$, the temperatures 
systematically increase. The heat current $I_{\rm h}$ caused by the electric current 
through the relationship of $I_{\rm h}=rI^2$ 
is controlled to keep 70, 80, 90, and 100 K at point(1). 
The most striking feature is that the magnitude of the heat current in the forward direction ($I_{\rm hf}$) 
is significantly larger than that in the reverse direction ($I_{\rm hr}$) showing the thermal rectification. 
The temperature difference between points (3) and (4) is completely explained by 
the thermal conductances of LCO and LSCO, 
which ensures that contact thermal resistance at the interface between LCO and LSCO 
is negligible. Note that contact thermal resistance between the rectifier and the cold head is also 
negligibly small. 

We evaluate this experimental results using Fourier's law. 
From Eq. (1), the temperature distribution of the thermal rectifier is derived as 
\begin{equation}
\begin{split}
&T(x)=T(0)-\int_{0}^{x}\frac{I_{\rm h}(\xi )}{\kappa [\xi ,T(\xi )]S(\xi )}{\rm d}\xi. \\
&I_{\rm h}(x)=I_{\rm h}(0)-\sigma \int_{0}^{x}l(\xi )\{T(\xi )^4 -T_{\rm C} ^4\}d\xi, 
\end{split}
\end{equation} 
where $I_{\rm h}$, $S$, $\sigma $, $l$, and $T_C$ represent heat current, sectional area, Stefan-Boltzmann constant, 
length of periphery of the sectional area $S$, and temperature of the cold head (40 K), respectively.
The radiation loss is properly treated in the second line of Eq. (2). 
It should be emphasized that $I_{\rm h}$(0) is only parameter for calculating the temperature 
distribution under a boundary condition decided by the experiments. 
Here, as the boundary condition, temperatures at point (1) ($x=0$ mm) and point (4) ($x=1.9$ mm) 
were set at 100 K and 40 K, respectively. As $\kappa $[$x$. $T$($x$)], $S$($x$), and $l$($x$), 
experimental data shown in Fig. 1 were used. 
As shown in Fig.2 (c), calculated temperature distribution of the rectifier is 
in good agreement with the experimental data, 
which shows that the heat current in the pyramid-shaped rectifier is consistently explained by Fourier's law.

\begin{figure}[t]
\begin{center}
\vspace*{0cm}
\includegraphics[width=5cm,clip]{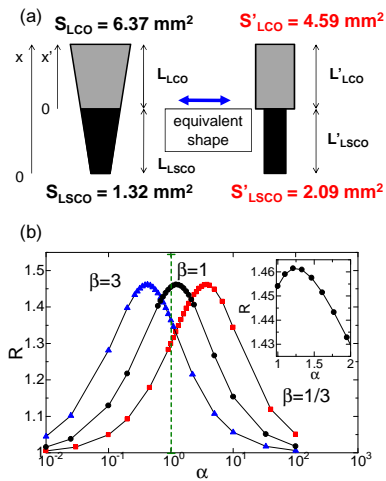}
\caption{(Color online) (a) Two equivalent thermal rectifiers which show same calculation result of $I_{\rm hf}$(0), 
$I_{\rm hr}$(0) and $R$ under same boundary condition. (b) Calculated $R$ as a function of $x$ and $y$. 
The inset shows the enlarged calculated result in the range of $1\leq x\leq 1.5$, $1.425\leq R\leq1.465 $.
}
\end{center}
\end{figure}

Fig. 3 (a) shows the calculated and experimental heat currents $I_{\rm hf}$(0) (LSCO-top configuration) 
and $I_{\rm hr}$(0) (LCO-top configuration) as a function of temperature difference 
$\Delta T\equiv$ $T$($x=$0 mm)-$T$($x=$1.9 mm). 
With increasing $\Delta T$, $I_{\rm h}$ monotonically increases, and experimental data are 
well reproduced by the calculation. 
As shown in Fig. 3 (b), the rectifying coefficient as a ratio of $I_{\rm hf}$(0) and $I_{\rm hf}$(0) 
is 1.35 at $\Delta T=$ 60 K, which is rather small compared with $R=1.43$ of the rectangular one.  

To address how $R$ depends on the thermal rectifier's shape, 
we perform numerical calculation of $R$ in a sample 
made of two different homogeneous materials with "a given shape". 
Fig. 4 (a) shows two equivalent thermal rectifiers which show same calculation results 
of $I_{\rm hf}$, $I_{\rm hr}$, interface temperature ($T_{\rm int}$) between LCO and LSCO, 
and $R$ under same boundary condition. 
To convert from a pyramid shape (a given shape) to an equivalent rectangular shape, 
a simple analytical relation is applied,  
\begin{equation}
\begin{split}
&\frac{L'}{S'}=\int_{0}^{L}\frac{dx}{S(x)} \\
\end{split}
\end{equation} 
where $L$($L'$), and $S$($S'$) represent a length and a sectional area of a given shape
(equivalent rectangular shape), respectively. 

Figure 4(b) shows the calculated $R$ as a function of $\alpha$ and $\beta$, where 
$\alpha \equiv \frac{S'_{\rm LSCO}}{S'_{\rm LCO}}$ ($S'_{\rm LSCO}$, $S'_{\rm LCO}$: 
sectional areas of LSCO and LCO with the equivalent rectangular shape), 
and $\beta \equiv \frac{L'_{\rm LSCO}}{L'_{\rm LCO}}$ ($L'_{\rm LSCO}$, $L'_{\rm LCO}$: 
lengths of LSCO and LCO with the equivalent rectangular shape). 
In the case of $\beta =1$, 
$R$ takes the maximum of 1.46 at $\alpha =1.2$. Thus, $R$ increases by 0.7$\%$ compared with 
a rectifier with $\alpha =$1 and $\beta =$1. 
Depending on $\beta $, $R$ profile shifts on $x$ axis, 
which is easily understood by thinking the thermal conductance of the thermal rectifier with 
rectangular shape. 
By applying this study to other materials, the thermal rectification may further enhance. 
We believe that this study will give a possible route to increase rectifying coefficient. 

We have investigated shape-dependent thermal rectification from both experimental and theoretical approaches. 
The cobalt oxide thermal rectifier with pyramid-like shape exhibits the rectifying coefficient of 1.35, 
which is rather small compared with that of rectangular-shaped thermal rectifier. 
The temperature distribution, heat current, and rectifying coefficient are consistently 
reproduced by Fourier's law. We have also shown that equivalent thermal rectifier with 
rectangular shape can be derived by a simple analytical relation, and found tiny enhanced 
rectifying coefficient at $\alpha =1.2$ and $\beta =$1. 
We believe that this study will give a possible route to increase rectifying coefficient. 

This study was partly supported by the Murata Science Foundation.

\end{document}